# Playing Bias: Epistemic Games for Interpreting Text-to-Image Models

Amalia Foka[0000-0001-7776-1308]

Department of Informatics and Telecommunications, University of Peloponnese, Tripoli, Greece
foka@uop.gr

**Abstract.** Text-to-image (T2I) systems do not merely render prompts; they circulate defaults for "what X looks like," shaping recognition through pose, camera, costume, props, and setting. While technical audits quantify disparities, they rarely cultivate the literacy needed to argue how images assemble stereotypes and confer authority. This paper translates a three-stage, humanities-informed framework into a set of epistemic games that make this cultural work visible and debatable: Interpret (art-historical close looking with AI outputs), Interrogate (counter-intent prompting as artistic exploration), and Expose (creative misuse through paired prompts with a constant motif and a BiasSlot rule, in which only the bias cue varies). In a pilot with undergraduate fine-arts students, the games shifted discussion from impressionistic reactions to concise, warranted hypotheses grounded in visual evidence, supported by lightweight classroom instruments that enabled reproducibility and structured collaboration. We argue that such game-based approaches foster critical literacy across educational contexts, equipping learners to recognize, discuss, and contest how T2I models encode bias.

**Keywords:** Text-to-Image, Generative AI, Bias, Critical Play.

## 1 Introduction

Text-to-image (T2I) systems do not just render prompts; they help establish everyday defaults for "what X looks like" through pose, camera, costume, props, and setting that persuade at a glance. Because these defaults travel through education, advertising, and policy materials, their semiotics have real downstream effects on aspiration and belonging. Existing evaluations can count disparities, yet they rarely teach us how to read these persuasive choices or argue about their effects. For example, across otherwise identical occupational prompts, models often portray men as equipped and agentic (tools, PPE, active posture) while staging women as aestheticized or passive (stylized hair, softened lighting, decorative framing) [1]. What is needed is a way to treat outputs as cultural artifacts and to apprentice non-specialists into making warranted, testable claims about them.

Art-historical methods offer exactly that footing. They provide a disciplined itinerary for reading images—moving from form (composition, pose, light) to iconography (motifs, symbols), to context (power, labor, belief)—and they turn looking into accountable interpretation.

Building on this logic, we previously articulated a three-stage, humanities-informed framework for evaluating T2I as cultural–technical systems: Interpret (close looking with AI outputs), Interrogate (counter-intent authoring as representational design), and Expose (paired prompts with a constant motif to reveal defaults) [2].

This paper argues that the framework should not remain a specialist protocol: it can and should be played. We translate its three stages into three epistemic games—Interpret, Interrogate, and Expose—that slow perception, structure debate, and render claims auditable, turning generative play into evidence. Together, these games provide a practical pedagogy that can travel across educational levels and settings, equipping learners and publics to recognize, discuss, and contest how T2I models assemble meaning.

We pursue two research questions framed in practice: (i) Does role-driven close looking improve participants' ability to form testable claims about model behavior? and (ii) Do collaboratively authored counter-intent prompts surface (and allow participants to challenge) stereotype assembly in models? Evidence from a fine-arts pilot indicates shifts in interpretive discipline and epistemic stance: participants moved from impressionistic reactions to concise, warranted hypotheses, and in paired comparisons they identified consistent representational defaults even under tightly controlled prompts.

Finally, we position this contribution within immersive learning research by treating immersion not only as a property of XR hardware but as an outcome of role-based, rule-governed collaborative inquiry: participants adopt epistemic roles, act under constrained interaction rules, and build shared claims from traceable evidence artifacts. This structure maps onto immersive learning designs—e.g., a virtual studio or gallery for embodied, side-by-side comparison of paired outputs and collaborative annotation—and yields actionable design primitives for immersive environments (roles, constrained actions, feedback, and inspectable artifacts) [3–5]. The connection to AI literacy is therefore mechanistic: the literacies we target depend on situated attention to visual cues, social negotiation of interpretations, and warranted justification, all of which immersive environments can strengthen through embodied comparison, social presence, and persistent annotations. We thus treat the epistemic games as transferable interaction patterns for immersive AI literacy that can be instantiated in XR or other immersive settings.

## 2 Background and Theoretical Framework

Machine learning systems do not only generate outputs; they also shape meaning-making by producing culturally legible signs that people interpret, circulate, and contest [6, 7]. Here, meaning-making refers to how generated outputs function as signs within human interpretive and social contexts, not to intrinsic machine semantics. Almost a decade of audits has shown that models routinely encode and amplify social disparities. For example, Buolamwini and Gebru's Gender Shades study documented stark intersectional accuracy gaps in face analysis systems [8], while Bolukbasi et al. demonstrated how word embeddings reproduce stereotypical analogies such as "man is to computer programmer as woman is to homemaker" [9]. These findings have spurred a repertoire of interventions: documentation practices in ML [10], fairness metrics and bias benchmarks [8], and mitigation strategies at the levels of data, prompt design, and architectural constraint [11, 12]. Such instruments quantify disparities and pressure pipelines toward accountability, but they rarely address the cultural work of images. These audits tell us what a model produced; they do not by themselves cultivate literacy to ask how those productions participate in stereotype assembly and social hierarchy.

Text-to-image (T2I) systems heighten these stakes because they generate the very medium through which contemporary culture organizes recognition and memory. Circulating images become working templates for categories—visual shortcuts that, through repetition, naturalize roles. When identical occupational prompts yield men posed as competent and women staged as decorative, the harm exceeds a mislabeled attribute: the model rehearses a visual grammar that shapes aspiration and belonging in public life. Studying bias in T2I therefore requires reading outputs as cultural artifacts, not merely as pixels or embeddings. As Crawford and Paglen argue, the training corpora that underwrite such systems must themselves be understood as visual archives, structured by politics, ideology, and historically situated regimes of vision [13].

Treating generated images as cultural artifacts also aligns with media theory accounts of software-mediated culture: contemporary visual production is increasingly conditioned by computational systems whose defaults and affordances shape what becomes visible, thinkable, and shareable [14]. From this perspective, T2I outputs can be approached as instances of software-driven cultural representation, co-produced by datasets, interfaces, and promptable generative procedures.

Art-historical methods provide a crucial entry point for this kind of analysis. Impett contends that the challenge lies not only in statistical disparities but in the "visual ideology" sedimented within training datasets, which function like photographic archives and demand an iconology of the dataset [15]. Rodríguez Ortega, in turn, describes T2I systems as instantiating a form of inverted ekphrasis: where art history traditionally translates images into words, generative models reverse the process by producing images from words, collapsing the classical tension between text and image into what she terms a post-semiotic vectorial regime [16]. Rose's framework for visual methodologies reinforces these perspectives by showing how visual culture must be analyzed across three sites—production, the image itself, and audience reception—and through technological, compositional, and social modalities [17]. Taken together, these insights resonate with the longue durée of art-historical practice, which proceeds from form to iconography to context. Bringing this apparatus to bear on T2I means recognizing that outputs are culturally coded, drawing on inherited repertoires of motifs and conventions. This motivates a framework that couples interpretive analysis with systematic experimentation to make those codes explicit.

Against this backdrop, we have previously adopted a three-stage framework that integrates art-historical close looking, artistic exploration, and critical prompt engineering to examine T2I models as cultural–technical systems [2]. First, canonical art functions as a calibrated lens for analysis: by staging AI outputs alongside well-theorized artworks, readers can move along a disciplined trajectory from form (composition, drapery, pose, light) to iconography (attributes, symbols, motifs) to context (historical conditions, power, labor). This Panofskian progression [18] transforms visual description into accountable claims about model behavior.

Second, artistic exploration operates as a mode of inquiry. Like Kehinde Wiley's reframing of canonical subjects, representational choices can be used to test how models encode cultural assumptions. In our framework this becomes representational design: set a counter-intent, align style and framing, vary one prompt slot at a time, and track shifts in authority or agency as evidence [2]. The goal is not merely to produce images but to generate testable hypotheses about when models resist or absorb representational change.

Third, creative misuse draws on critical theory to probe how T2I systems reproduce hierarchies. Prompts differ by a single bias-bearing cue—gender, race, class, age, disability—while style and composition remain fixed. A constant symbolic motif (e.g., crown, flower, safety gear) deliberately unsettles priors. Read through feminist, postcolonial, or disability studies, such pairings reveal how authority, passivity, modernity, or tradition are visually encoded. Studies of labor roles show that seemingly "identical" prompts still yield stereotyped contrasts [1]. Creative misuse thus joins controlled variation with critical theory, showing how models perpetuate or aestheticize long-standing ideologies.

Together, these strands—art-historical analysis, artistic exploration, and creative misuse—form a framework for studying T2I models as cultural–technical systems. They also pose a pedagogical challenge: generative images shape how societies imagine work, identity, and power. If only technical experts can interpret them, their cultural effects remain opaque. Teaching students and publics these practices equips them to spot stereotypes, question defaults, and contest how AI images are used. Such literacy is both academically valuable and socially urgent, fostering critical awareness in everyday encounters with generative media.

To meet this challenge without trivializing ethics, we treat play as a learning stance characterized by active exploration, low-stakes experimentation, and iterative hypothesis testing under constraints. Work on playful learning and guided play shows that structured play can sustain engagement while supporting disciplined inquiry when goals, feedback, and reflection are built into activity design [19, 20]. This makes play well-suited to bias interpretation: it licenses systematic "what-if" variations while the scaffolds channel attention toward evidence, comparison, and justification. In our design, roles, constrained prompt actions, and inspectable artifacts function as guided-play supports, making interpretive claims auditable rather than impressionistic; a digital game implementation would allow these constraints and feedback loops to be enacted consistently at scale.

Games provide a uniquely effective vehicle for this apprenticeship. As rule-based systems, they afford meaningful play through structured goals, feedback, and consequences [21]. They also persuade through rules—what Bogost terms procedural rhetoric—demonstrating how systems encode values and arguments [22]. Designed as critical play, games deliberately subvert conventions to open spaces of reflection and social change [23]. In educational research, epistemic games have shown how roles and structured inquiry can scaffold learners into disciplinary ways of seeing and reasoning [24]. Building on media literacy and participatory design traditions, these methods also foreground co-creation: learners do not simply interpret media but actively remix, produce, and reconfigure it, developing interpretive, ethical, and productive agency [25–27].

Transposed to the domain of T2I, games transform critical image inquiry into a participatory and accountable process: they slow perception, structure debate, and turn abstract theories of bias into lived, collaborative exploration. In this way, game-based methods make visible the cultural logics of generative AI and democratize who gets to interpret them, cultivating interpretive, ethical, and productive literacies.

**Table 1.** Epistemic games: phases and instruments.

| Phase | Artifact(s) | Scaffold(s) | Core Instrument(s) | Claim Type |
|---|---|---|---|---|
| Interpret | Evidence Grid; Exit Memo | Evidence Grid; Exit Memo | Roles & Ledger | Observation → Interpretation → Hypothesis |
| Interrogate | Prompt Row; Semiotic Palette; Exit Memo | Prompt Row; Semiotic Palette; Counter-Intent Card | Evidence Grid + Semiotic Palette + Prompt Row | Hypotheses on how models handle counter-intents |
| Expose | Paired Prompts (BiasSlot); Findings Track; Exit Memo | Paired Prompts; Findings Track | BiasSlot + Constant Motif + Findings Track | Hypotheses on representational defaults and encoded bias |

## 3 Methods: Epistemic Games for Critical Image Inquiry

The three phases—Interpret, Interrogate, and Expose—translate art-historical analysis, artistic exploration, and creative misuse into structured classroom games. Table 1 summarizes how each phase aligns its *artifacts* (what learners produce), *scaffolds* (supports that guide attention and discussion), and *core instruments* (the rule mechanisms that constrain action) with the type of claim it develops. For example, in *Interrogate* the Prompt Row constrains variation across generations, while the Semiotic Palette specifies the visual cues that count as evidence for or against a counter-intent. We elaborate the three phases as game rounds (Rounds 1–3) in the sections that follow.

### 3.1 Phase 1 — Interpret: Art-Historical Close Looking with AI Outputs

Phase 1 is a structured game of collaborative close looking that trains students to treat AI-generated images as cultural artifacts by reading them alongside a canonical artwork. The design follows an art-historical sequence: disciplined description of form, interpretation of iconographic elements, and contextualization within broader cultural and ethical frameworks. In a single compact session, students learn to (i) describe what they see with precision, (ii) connect formal features to symbolic and cultural meanings, and (iii) state a concise, testable hypothesis about how and why a model produced particular results.

The game begins with a short instructor primer that models the move from form to meaning on the chosen artwork and introduces a small, controlled set of AI outputs generated from prompts. The aim is not to front-load interpretation but to establish a shared vocabulary and a transparent basis for comparison. Typically two or three outputs are supplied—for example, one from a formal-descriptive prompt (composition/form without naming titles), one from a symbolism-led prompt (iconographic content over style), and optionally a controlled-variation prompt (isolating a single change in setting or period). This limited set keeps attention on interpretation rather than production (see Figure 1).

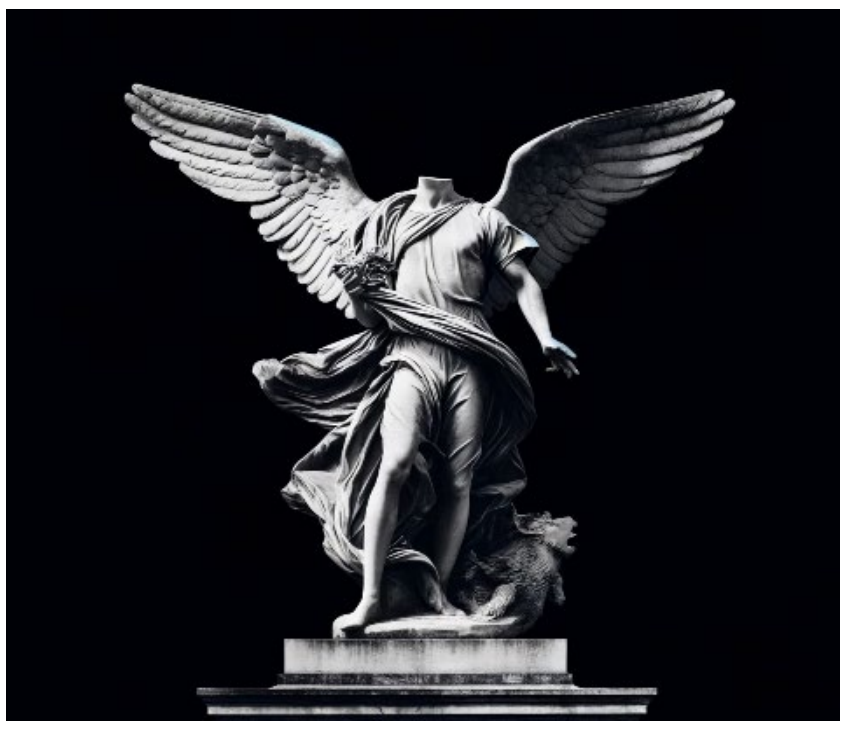

(a) DALL·E - Nike, symbolism-led.

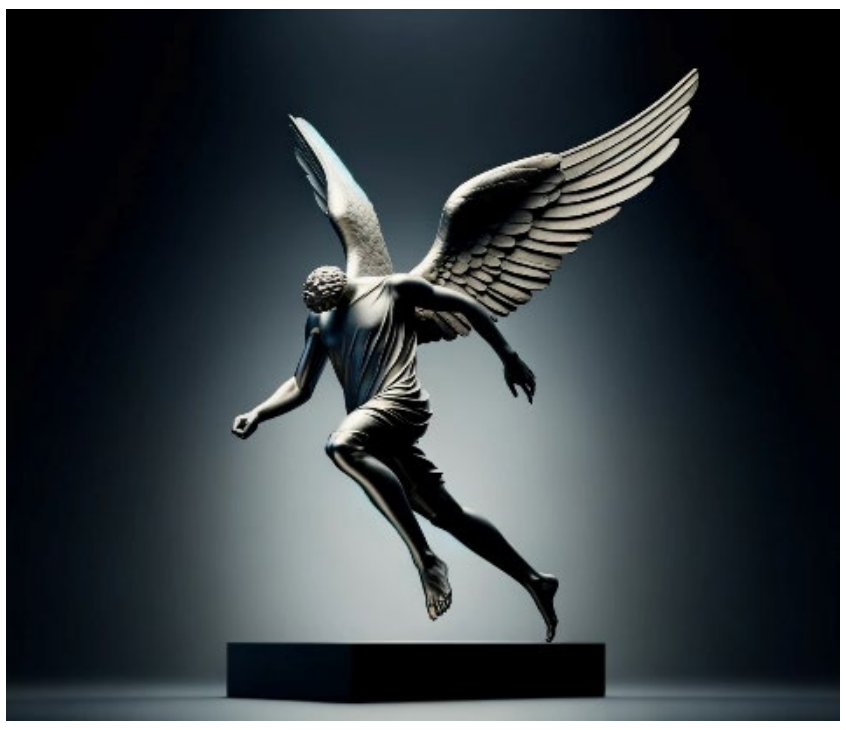

(b) DALL·E - Nike, modern variation.

**Fig. 1.** DALL·E exemplar outputs for the Winged Victory of Samothrace: symbolism-led vs. controlled-variation prompts. Images: A. Foka.

The core of the game is the evidence sweep, where teams transform looking into analyzable data. The instructor poses a broad interpretive question—such as how gender, class, or power is represented in a given role—so that all teams share a common horizon of inquiry. Within each team, the *Curator* reformulates this into a more specific angle that can guide comparison across images and keeps the group focused during discussion. The *Formal Analyst* leads on compositional and material features—pose, framing, scale, light—ensuring claims remain grounded in what is visible. The *Iconographer* identifies symbols, attributes, and recurring motifs, linking them to art-historical traditions. The *Contextualist* highlights cultural and ethical stakes, such as gendered labor, class coding, or colonial tropes, keeping the analysis socially situated. The *Archivist* records observations and claims verbatim so that reasoning remains transparent and can be revisited later. Roles may be merged in smaller teams (e.g., combining *Formal Analyst* and *Iconographer*) to maintain flow without overloading participants.

Scoring is diagnostic rather than competitive: it paces attention and rewards careful reasoning. Teams earn *Observation* points (+1) for concrete, non-inferential details (e.g., "the figure's hands are empty; the gaze is directed upward"); *Interpretation* points (+2) when those details are linked to symbolic or cultural meaning (e.g., "the elevated gaze connotes spiritual authority"); and *Hypothesis* points (+2) for concise, testable claims about model behavior (e.g., “the AI tends to associate power with male figures”). To keep the round moving, each team can earn a maximum of ten points in total across all images discussed. Once this cap is reached, scoring shifts from accumulating points to sharpening the clarity and warrant of claims.

Each session ends with a brief micro-debrief in which teams present their strongest observation–interpretation pair and one bias hypothesis tied to the guiding angle. This is consolidated in a 150–200 word exit memo that cites visual evidence from the canonical work and at least two AI outputs, states the claim in one or two sentences, and proposes a follow-up test. For example, a team hypothesizing that the model links power with male figures might suggest rerunning the prompt with only gender varied to see whether authority markers (pose, scale, attire) change accordingly.

### 3.2 Phase 2 — Interrogate: Counter-Intent Prompting as Artistic Exploration

In Phase 2, artistic practice becomes a *collaborative authoring game*: students co-author with the model, moving from the interpretive work of Phase 1 to structured experimentation. The activity adapts methods of artistic exploration—iterative, research-driven, and experimental—into a classroom routine where inquiry, collaboration, and play drive the design of prompts. Prompt engineering is treated as narrative construction: every descriptor is grounded in evidence and chosen for its representational consequences.

The game has three aims: (i) to help students recognize how stereotypes around themes such as “housekeeper,” “nurse,” or “refugee” are assembled across culture; (ii) to translate those insights into counter-intent prompts that avoid direct naming and instead rely on indirect, counter-stereotypical description; and (iii) to show—through controlled single-slot variation and reflection—how models reproduce or resist these patterns. We treat the model not as a neutral tool but as a collaborator whose tendencies must be tested. To foreground agency, we adopt Kehinde Wiley’s aesthetic—where ornament and monumental scale confer dignity—as a counter-stereotypical semiotics for reframing roles [28]. Figure 2 illustrates this approach with results from an earlier study on the “housekeeper” theme, using flowers and vibrant color in Wiley’s empowering mode.

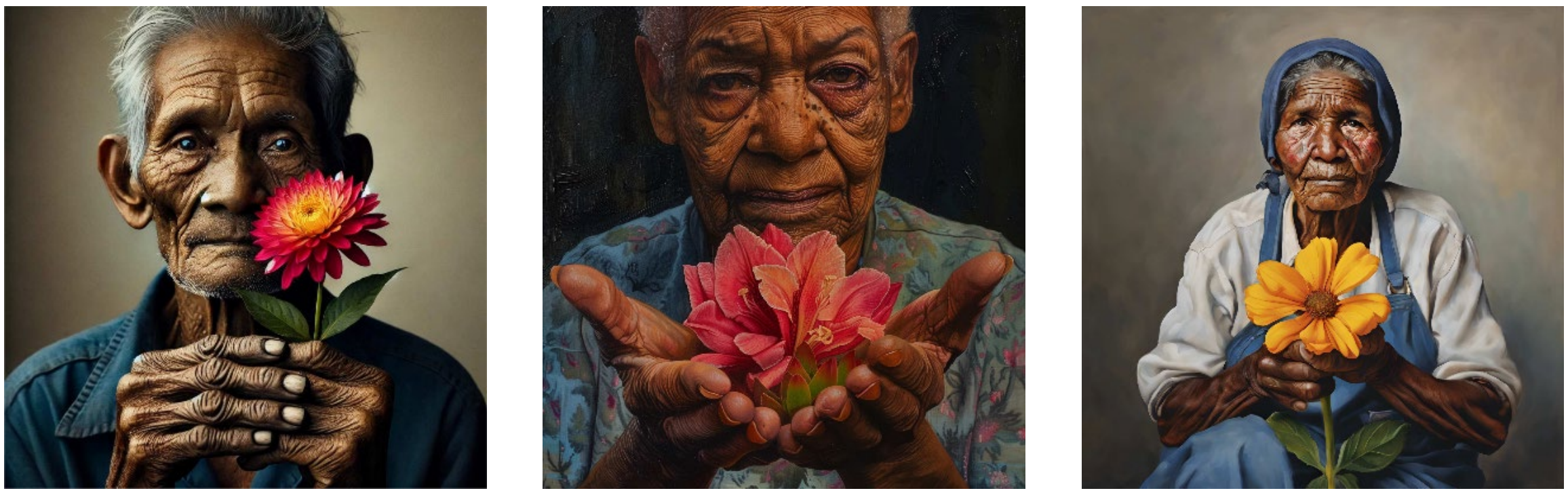

(a) DALL·E (b) Midjourney (c) Stable Diffusion

**Fig. 2.** Wiley-inspired domestic labor prompt outputs across three models. Images: A. Foka.

The game is played on a structured *board*, where each round fills a different section:

**Round 1: Evidence Collection**. Students begin by filling the *Evidence Grid* (a multi-domain database of stereotypes). Each member contributes at least one entry until the grid contains a minimum of six examples drawn from at least three domains: visual culture (film, advertising, photography), literature (journalism, fiction, testimony), iconography (recurring motifs, postures, objects), language (synonyms, euphemisms, slurs), or contextual frames (historical, cultural, social). Each entry is dated and culturally located to ensure specificity. The team then synthesizes a short *Stereotype Card* (2–3 sentences) summarizing typical representations; this serves as the baseline to be challenged in later rounds.

**Round 2: Intent Framing**. Students shift to what they want to foreground instead of stereotypes. They fill the *Semiotic Palette* (a set of Narrative Intents such as agency or care) with two to three *Narrative Intents*—values such as agency, solidarity, expertise, care, or joy—grounded in Evidence Grid entries and/or explicit ethical reasoning. From this pool, they draft a 2–3 sentence *Counter-Intent Card* that states their representational goals, ensuring authoring is guided rather than arbitrary.

**Round 3: Prompt Authoring**. Teams use the *Prompt Row* (four slots: Subject, Action, Style, Composition) with mandatory fields: *Subject (indirect)*—attributes/qualities rather than labels; *Action/Scene*—activity or setting; *Style/Aesthetic*—a mode that reinforces the counter-intent (portrait, documentary realism, allegory, poster); and *Composition/Lighting*—choices that redistribute authority and presence (e.g., eye-level vantage, central placement, dramatic lighting). Every descriptor must be traceable to the Evidence Grid or the Counter-Intent Card. Slots are assembled into a complete prompt and logged verbatim. When platforms exposed seed values and generation settings (e.g., steps, guidance), we fixed seeds within each variation set to reduce stochastic drift and rotated seeds across sets to check that findings were not seed-specific.

**Round 4: Testing and Findings**. Each team generates one *direct-label* image that names the subject explicitly (e.g., "housekeeper" or "refugee") as a baseline. They then run their counter-intent prompt and generate at least two further images by varying only one slot at a time (e.g., changing *Action/Scene* while holding all else constant), yielding a minimum set of four images (baseline, counter-intent, two single-slot variations). Prompts, outputs, and—where possible—model metadata (model/version, seed, sampler) are logged with unique identifiers for auditability. Teams note recurring features, deviations, and moments of resistance, and complete the *Findings Track* (bias hypothesis, one concise piece of visual evidence, and a next-step test). These are consolidated in a required 150–200 word exit memo stating the claim, citing visual evidence, and proposing a follow-up variation.

Scoring is diagnostic rather than competitive, with a maximum of ten points distributed across stages to keep attention balanced. Credit is awarded for completing a varied Evidence Grid, drafting a Counter-Intent Card that explains how selected values resist stereotypes, filling all Prompt Row slots with justified descriptors, producing a comparison set (direct-label baseline plus at least two single-slot variations), and concluding with a Findings Track that states a concise bias hypothesis with supporting evidence. Points therefore register progress through research, design, iteration, and reflection, ensuring no single step dominates and preserving the rhythm of artistic exploration.

By the end of this phase, students have transformed research into representational design, reasoned about AI outputs as culturally conditioned responses, and produced testable claims about bias and aesthetics. Framing prompt engineering as collaborative authorship keeps the process playful yet rigorous: evidence, intent, and ethical reflection remain entangled at every stage.

### 3.3 Phase 3 — Expose: Creative Misuse via Paired Prompts and a Constant Motif

In this phase, prompt engineering is treated as *critical play*. Students practice "creative misuse," deliberately bending expected uses of the model to expose hidden tendencies. The aim is not aesthetic polish but disruption: pairing a masculine role with a delicate symbol or inserting an allegorical motif that unsettles conventional settings. By misusing the model, students prompt it to reveal how it reproduces, reinforces, or resists representational bias. Teams work as a unit in a simple *propose–challenge–lock* cadence that keeps decisions accountable and logged.

The game proceeds in seven structured stages:

**1. Selecting Bias and Theory.** The instructor identifies a bias to interrogate—gender, race, class, age, disability, migration, religion, or sexuality—and introduces a critical lens for the exercise. A short lecture provides cultural and historical grounding so students can connect surface patterns to structures of power (e.g., gender bias via feminist theory; postcolonial bias via Orientalism).

**2. Choosing the Theme.** The theme is a culturally saturated role, identity, or situation aligned with the chosen bias and recognizable through common tropes. For example, age bias → "tech entrepreneur"; class bias → "CEO" or "politician." The theme is then locked so all subsequent steps build on it consistently.

**3. Defining Axes of Comparison.** The team defines at least three axes phrased as visual oppositions (e.g., authority vs. subordination; agency vs. passivity; professionalism vs. aestheticization). Axes are derived by asking: *How is this group typically represented? What qualities are erased or distorted?* Once written, axes are locked and displayed on the worksheet.

**4. Introducing a Symbolic Motif.** The team introduces a symbolic motif (a constant, unsettling allegorical element) and *locks it* so it remains fixed across prompts; this makes its effects traceable. *Motif heuristic: select motifs coded with opposing values (e.g., softness vs. toughness) or with allegorical resonance; reject motifs that collapse into mere ornament in pilots.* Motifs inject alternative values (e.g., flowers, crowns, dance, mythic backdrops), expanding the representational space: a crown can reframe manual labor as authority, while a flower can interrupt martial toughness with fragility.

**5. BiasSlot Rule.** In each prompt pair, exactly one bias-bearing cue changes (e.g., gender, age, ethnicity), while all other descriptors and the constant motif remain fixed; this isolates the effect of the chosen variable.

**6. Prompt Construction.** Prompts are authored via a structured template: *Subject*, *Action/Scene*, *Style/Aesthetic*, *Composition/Lighting*, plus the constant *Symbolic Motif*. Teams apply the *BiasSlot* rule so only one cue differs between paired prompts; everything else remains locked. Each team produces three paired prompts (six total). All text and, where possible, model parameters (model/version, seed) are logged verbatim for auditability.

**7. Generate, Compare, and Reflect.** Students generate the six images and compare them pairwise along the predefined axes. They ask: does authority shift when gender changes? Does agency increase or diminish with age? Does the motif counter stereotypes or get absorbed as ornament? Observations are tied to specific cues—pose, clothing, tool use, background—and logged. For example, Figure 3 shows a construction–ballet diptych in which gender varies under the BiasSlot rule while the ballet motif remains constant. Reading the pair along axes of authority vs. aestheticization and safety vs. spectacle reveals how the model codes labor and embodiment differently by gender.

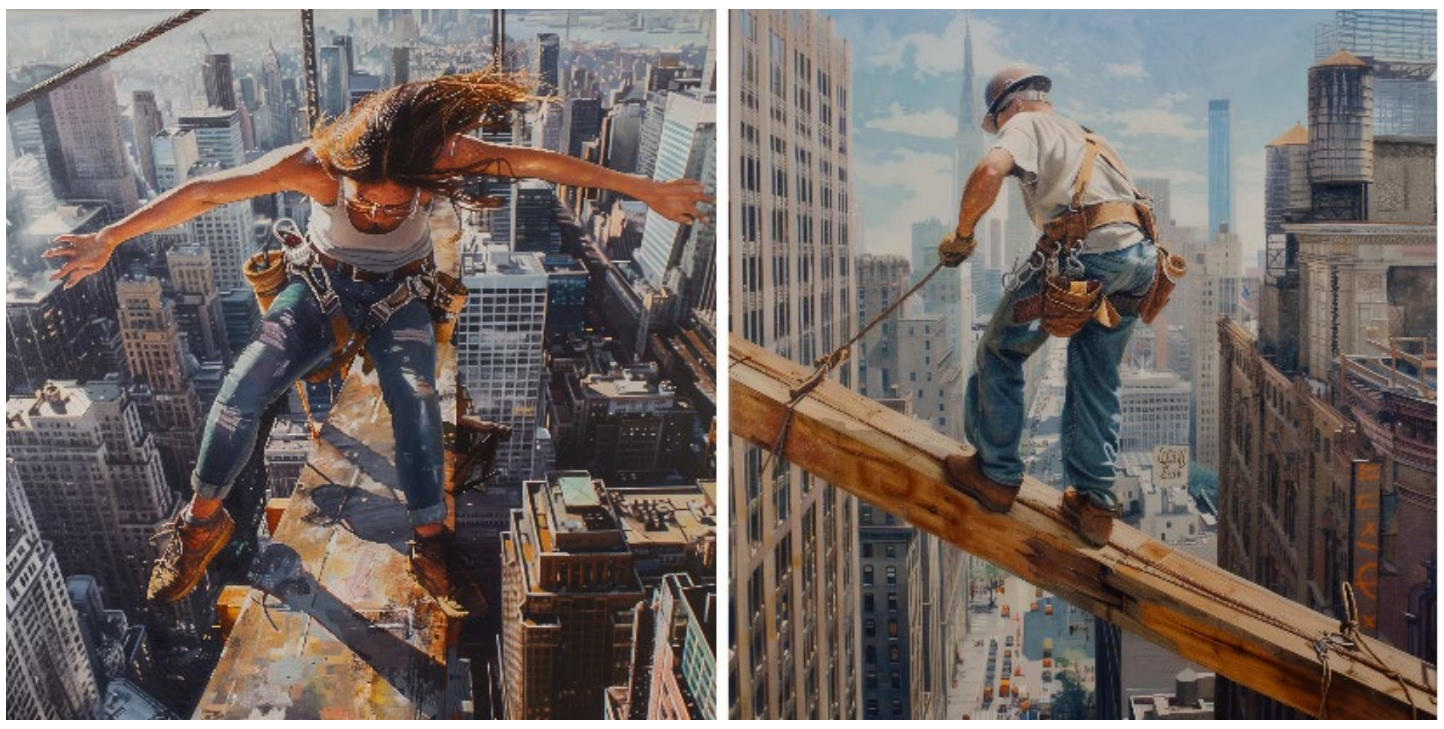

**Fig. 3.** Construction ballet diptych with BiasSlot = gender; ballet motif held constant. Images: A. Foka.

Scoring in this phase is diagnostic rather than competitive: each team can earn a maximum of ten points across the round. Credit is awarded for formulating paired prompts that isolate a single bias cue under the BiasSlot rule, documenting outputs with a constant motif, and completing a Findings Track that states a clear bias hypothesis supported by visual evidence. As in earlier phases, the cap guides attention without turning play into competition.

The round closes with a 150–200 word exit memo stating the team's strongest observation–interpretation link, a testable bias hypothesis, and one follow-up prompt variation. A small progress score marks each stage completed—axes defined, motif justified, BiasSlot applied, prompts constructed, outputs generated, memo submitted—so that research, design, iteration, and reflection remain balanced. In this way, creative misuse becomes a systematic classroom experiment rather than random provocation.

## 4 Pilot Study in a Fine-Arts Studio Course

The three games were implemented in an undergraduate studio course on *Artificial Intelligence and Art* taught by the author at the School of Fine Arts, University of Ioannina (Greece), with 15 enrolled students. Students were familiar with art-historical methods, visual analysis, and artistic exploration, but had little or no background in coding. The intervention ran over six 3-hour studio classes (two sessions per game). Teams were configured per game (and could be re-formed between games due to attendance and platform constraints), yielding 11 *team instances* across the pilot (4 in Phase 1, 3 in Phase 2, 4 in Phase 3). We collected one short exit memo per team at the end of each session, resulting in 22 memos in total, alongside the completed game artifacts (Evidence Grids, Cards, Prompt Rows, Findings Tracks) and generation logs (prompts, model/version, seeds where available). Pilot claims are based on descriptive memo statistics (submission rate, mean length, and presence of ≥2 observation→interpretation links) and a qualitative synthesis of recurring patterns in memos and artifacts; the pilot was not designed as a controlled experimental evaluation. This context shaped both the design of the pilot and the way students responded to the activities. The games drew conceptually on earlier research–creation projects that informed the classroom design—Phase 1 [29], Phase 2 [2], and Phase 3 [1]—which had experimented with art-historical comparison, collaborative authoring, and feminist critiques of generative AI. For the classroom, these projects were reworked into playable formats that could foster critical reflection and collaborative inquiry without assuming technical expertise.

Across all three phases, gender bias served as the primary lens, though issues of class and race also surfaced. In Phase 1, canonical works such as the *Winged Victory of Samothrace* and Judy Chicago's *The Dinner Party* were used to scaffold analysis. Students were given a small set of instructor-supplied exemplars (see Figure 1). At first, students found it difficult to remain within the structured "evidence sweep" of the game, preferring to generate many images and discuss them informally. As the round progressed, however, the value of the method became clear: the turn-based roles and diagnostic scoring slowed down impulsive exploration and required students to ground their observations, interpretations, and hypotheses in concrete visual evidence. This shift from casual talk to accountable comparison allowed bias hypotheses to emerge directly from what was seen—for example, "the model tends to associate power with male figures" or "sexual symbolism is avoided even when implied."

In Phase 2, the theme of the housekeeper anchored the collaborative authoring game. Students populated an Evidence Grid with examples from advertising, film, and stock photography, tracing how domestic labor has been feminized and racialized in European contexts. They then drafted Counter-Intent Cards inspired by Kehinde Wiley's portraiture, using symbolic ornament and vibrant color as counter-motifs (see Figure 2 for the example that informed this design). Wiley's style was chosen because his large-scale, ornamented portraits visually confer dignity and empowerment, providing a counterpoint to stereotypes of domestic labor. Cross-model comparison showed how counter-intents were absorbed or resisted: DALL·E and Midjourney depicted both genders, though skewed toward women of color, while Stable Diffusion produced only female figures. All three models emphasized age, influenced by the phrase "lifetime of domestic labor." While some images conveyed resilience, the vibrancy and empowerment characteristic of Wiley's work were diluted. The structured Prompt Row—though initially frustrating for students—proved essential: small, controlled variations revealed how models reasserted stereotypes, demonstrating that such outcomes are not incidental glitches but structural tendencies of generative systems.

In Phase 3, a prior research–creation methodology developed by the author for exploring gender bias in depictions of labor was adapted into a classroom setting, with the carpenter chosen as the theme. Construction work, as a paradigmatically masculine-coded profession, offered a clear test-bed for examining how T2I systems differentiate gender under otherwise identical conditions. Students defined axes such as authority versus passivity, professionalism versus aestheticization, and safety versus spectacle, and introduced ballet as a constant symbolic motif to unsettle defaults. Prompts were authored under a strict BiasSlot rule (only gender varied; all other descriptors and the motif remained fixed). The guiding example for this phase was the construction–ballet diptych shown in Figure 3, where the constant motif of ballet highlighted how the model encoded gender differently along the chosen axes. The resulting pairs consistently contrasted competent, well-equipped male carpenters with aestheticized female counterparts posed decoratively with flowing hair or dramatic lighting—images closely resembling advertising tropes or over-beautified stock photography. Exit memos then articulate bias hypotheses in testable form, linking observations back to cultural patterns.

Several implementation issues surfaced across phases. Students initially struggled to work in teams and to understand the value of the assigned roles, especially in Phase 1, but by Phases 2 and 3 they were more comfortable with the cadence of propose–challenge–lock decisions. Technical hurdles also appeared. While this pilot ran on paid versions of the platforms, instructors working with free plans will have to navigate current limitations, particularly caps on generated images; these constraints may force teams to be selective in experimentation. For classes where account creation is constrained (e.g., age/GDPR), we recommend instructor-run generation or a pre-generated set from a single reliable model to ensure equity across teams. For sensitive identity prompts, generation should remain instructor-run to maintain safety oversight and prevent exposure to harmful content. Another recurrent challenge was that students could readily observe biases but often struggled to move from description to explanation; a short, instructor-supplied glossary linking critical terms to visual cues (e.g., "male gaze" → camera angle and gaze direction; "ornamentalization" → pose and prop treatment) helped bridge this gap. Furthermore, since model training data are not accessible—and even if they were, their scale makes them unmanageable—students were at times directed to conduct image searches online (e.g., stock photography) with relevant keywords. These searches provided hints about the kinds of imagery that dominate online sources, which are known to feed into training data, and helped students formulate more grounded explanations of the biases they observed.

Despite these challenges, student engagement appeared high, inferred from regular memo submissions, sustained time-on-task during rounds, and instructor observation. Many enjoyed the game format, finding it playful yet rigorous, and were struck by how quickly biases became visible once prompts were systematically varied. Some initially believed that gender stereotypes had weakened in contemporary culture, but the outcomes led them to reconsider how such stereotypes persist in visual production, especially through training data drawn from commercial imagery. Their attitudes toward AI image generators were complex: as fine-arts students they were skeptical of the models' creative potential, but they came to appreciate their technical capacities and, more importantly, to see them as sites for critical inquiry rather than as neutral tools. Students from other disciplinary backgrounds may approach the models differently, often assuming they are almost fully creative; for them, the game framework could provide a valuable corrective, clarifying both the potential and the limits of machine creativity.

The progression across phases was clear. Phase 1 cultivated interpretive discipline and hypothesis building; Phase 2 taught students how to convert critique into counter-intentional design; Phase 3 extended this into creative misuse, using symbolic motifs and controlled variation to expose entrenched representational logics. If the sequence were repeated, adjustments might include streamlining the role system in Phase 1 and allocating more time for reflective writing at the end of each phase. For instructors adopting the framework in different contexts, two points are especially important: first, students without prior training in art history or visual culture may need additional scaffolding to analyze images systematically; and second, careful attention should be paid to platform and model choice, including free–paid options and their limitations, given the pace of technological change. As with any classroom pilot, findings reflect the specific models and prompts used; generalization is bound. Overall, the pilot showed that epistemic games can be successfully integrated into an art-studio environment, transforming playful experimentation into structured critical reflection, and enabling students to see generative AI as both a technical system and a cultural agent.

## 5 Findings & Discussion

Across the three phases (Interpret, Interrogate, Expose), the games proved workable in studio conditions and diagnostic of how T2I systems organize visual meaning. Drawing on the exit memos, logged artifacts, and generation records, we observed a shift from impressionistic talk to accountable comparison: students increasingly formulated short, testable hypotheses grounded in specific cues (pose, camera angle, apparel, tool handling, background). Participation was sustained across the six sessions: the 11 team instances produced 22 exit memos (mean $\approx$173 words), each containing $\geq$2 observation$\rightarrow$interpretation links.

In *Interpret*, the role structure and point ledger slowed perception and reduced speculative leaps; learners consistently tied claims to what was visible, and hypotheses emerged directly from contrasts among the supplied outputs. This addresses **RQ1**: role structure plus diagnostic scoring increased the rate and quality of *testable claims*, evidenced by sustained memo production with explicit links from observations to interpretations and a representative coded example tying claims to features such as camera height, stance, PPE, and gaze.

In *Interrogate*, the Evidence Grid → Counter-Intent → Prompt Row pipeline translated critique into representational design: single-slot variations revealed where counter-intents were absorbed, resisted, or neutralised by the model, and teams learned to treat such failures as evidence rather than glitches [2]. In *Expose*, paired prompts under a clear BiasSlot rule, together with a constant symbolic motif, made defaults legible: under otherwise identical conditions, labor-role pairs repeatedly contrasted competent, well-equipped figures with aestheticized counterparts, echoing commercial stock tropes and yielding succinct bias hypotheses that students could justify with concrete features [1]. Taken together, these results answer **RQ2**: counter-intent prompting with single-slot variation reliably surfaced stereotype absorption/resistance, with corroboration from the cross-model "housekeeper" theme and the construction–ballet diptych.

Pedagogically, the scaffolds functioned as intended: roles distributed cognitive work; diagnostic scoring paced attention; exit memos consolidated warrants. This aligns with findings on epistemic games and the capacity of rule-based activities to structure meaningful inquiry and argument [21–24]. Two lightweight supports were pivotal in practice: a mini-glossary mapping theoretical terms to visual cues (e.g., "male gaze" → camera angle/gaze direction; "ornamentalisation" → pose/prop treatment), and systematic logging (model/version, seed, prompt ID), which made comparisons auditable and discussions replicable [29]. Taken together, the activities complement technical audits by surfacing semiotic operations through which images confer authority, agency, or ornament—features that aggregate scores do not capture—and they do so in a form accessible to non-specialists.

Coding was descriptive and conducted by a single rater, with no inter-rater reliability; accordingly, we present these findings as *illustrative* rather than confirmatory and plan a second-coder pass with IRR in future iterations. Threats to validity include model drift (time/version), addressed by logging and to be addressed through cross-version replications; seed sensitivity, addressed (where available) by fixed seeds within variation sets and planned multi-seed sweeps; cohort specificity (fine-arts students), to be tested via replications in non-art disciplines (e.g., education, media/communication, computer science) and mixed cohorts, and through public-facing workshop deployments to assess transfer beyond studio-based visual literacy; and motif dependence, mitigated by a stated motif heuristic and pilot screening, with planned ablations comparing multiple motifs per theme.

## 6 Conclusions & Future Work

This study sets out to close a gap between what technical audits can count and what images actually *do* in culture. By translating a three-stage framework into playable classroom practices, we showed how learners can turn generative outputs into accountable inquiry: first by looking closely and making warranted claims; then by designing counter-intent prompts and testing them systematically; and finally, by using paired comparisons to surface representational defaults. In effect, the games make visible meaning-making operations—pose, framing, attire, tools, setting—that aggregate scores leave offstage, and they do so through structured, guided play: low-stakes exploration and iterative hypothesis testing under constraints, where reflection and evidence keep play ethically serious rather than trivial [19–21, 23].

Because the method is lightweight and role-based, it can travel across settings—from fine arts and media studies to computer-science ethics and secondary digital literacy—and across lenses (age, disability, migration) and languages. At the same time, our pilot cohort was fine-arts students; broader disciplinary testing is therefore a priority before making strong generalizability claims. For instructors, a simple throughline proved reliable: begin with close looking; require a Counter-Intent Card; lock a constant motif; enforce a BiasSlot rule; log model/version/seed and prompts; and keep scoring diagnostic rather than competitive. These moves slow perception, structure debate, and make claims auditable.

Looking ahead, we will broaden both reach and rigor through multi-class, multi-institution replications that include non-art cohorts (e.g., education, media/communication, computer science) and mixed groups, alongside public workshops, to test transfer and identify discipline-specific scaffolding needs. We will pair these deployments with pre/post measures of interpretive and ethical literacies, cross-model replications (open and proprietary) with version tracking, and extensions to additional modalities (e.g., text-to-video, multimodal agents) to examine temporal stereotypes. To directly address *model drift* and *seed sensitivity*, we will run cross-version, multi-seed robustness checks and release full provenance logs; to test *motif dependence*, we will pre-register motif sets and conduct motif ablations. A complementary direction is to integrate interpretive findings with quantitative audits so that metrics and semiotic evidence inform one another. We also plan to release an open toolkit—role cards, Evidence Grid, Semiotic Palette, BiasSlot templates, and rubrics—to ease adoption.

We also situate this work within immersive learning by treating immersion not only as a property of XR hardware but as an outcome of structured interaction: epistemic role-play, constrained action, feedback, and shared inspectable artifacts. These design primitives map naturally onto immersive environments—e.g., virtual studios or galleries where outputs can be inspected at scale, compared across embodied viewpoints, and annotated collaboratively—supporting social presence in debate and persistent traceable evidence for AI literacy [3–5]. In this sense, our contribution is a set of transferable interaction patterns for immersive AI literacy that can be instantiated in XR or other immersive platforms.

A promising next step is the *digital game* implementation of these activities. Embedding the flows, rules, and logging directly in an interface would operationalize the pedagogy of guided play—enforcing constraints such as the BiasSlot and constant motif, prompting reflection, and automating provenance (prompts, seeds, versions)—while providing live visualizations of Observation–Interpretation–Hypothesis chains. Such a platform could support remote and asynchronous participation, enable cross-institution comparison, and generate rich play traces for research, without replacing the discursive richness of the classroom.

To confront AI bias, we must not only *measure* it, but also *teach* it, *discuss* it, and—crucially—*play* with it. Doing so equips students and citizens to read T2I systems as both technical artefacts and cultural mirrors.

## 7 References

1. Foka, A.: She Works, He Works: Critical Explorations of Gender in AI-Generated Representations. In: *Proceedings of the International Symposium on Electronic Art (ISEA 2025), Seoul, Republic of Korea*, pp. 630–637. Art Center Nabi and Institute for Culture and Arts, Seoul National University, Seoul (2025). doi:10.23362/KOEN2025.07.25.1.082.
2. Foka, A.: A Framework for Critical Evaluation of Text-to-Image Models: Integrating Art Historical Analysis, Artistic Exploration, and Critical Prompt Engineering. In: Del Bue, A., Canton, C., Pont-Tuset, J., Tommasi, T. (eds.) *Computer Vision – ECCV 2024 Workshops*, LNCS 15644, pp. 128–143. Springer (2025). doi:10.1007/978-3-031-92089-9_9.
3. Dede, C.: Immersive interfaces for engagement and learning. *Science* 323(5910), 66–69 (2009). doi:10.1126/science.1167311.
4. Dalgarno, B., Lee, M.J.W.: What are the learning affordances of 3-D virtual environments? *British Journal of Educational Technology* 41(1), 10–32 (2010). doi:10.1111/j.1467-8535.2009.01038.x.
5. Makransky, G., Petersen, G.B.L.: The cognitive affective model of immersive learning (CAMIL): A theoretical research-based model of learning in immersive virtual reality. *Educational Psychology Review* 33, 937–958 (2021). doi:10.1007/s10648-020-09586-2.
6. Gillespie, T.: The relevance of algorithms. In: Gillespie, T., Boczkowski, P.J., Foot, K.A. (eds.) *Media Technologies: Essays on Communication, Materiality, and Society*. MIT Press, Cambridge, MA (2014).
7. Seaver, N.: Captivating algorithms: Recommender systems as traps. *Journal of Material Culture* 24, 421–436 (2019).
8. Buolamwini, J., Gebru, T.: Gender shades: Intersectional accuracy disparities in commercial gender classification. In: Friedler, S.A., Wilson, C. (eds.) *Proceedings of the 1st Conference on Fairness, Accountability and Transparency (FAccT 2018)*, pp. 77–91. PMLR (2018).

9. Bolukbasi, T., Chang, K.-W., Zou, J.Y., Saligrama, V., Kalai, A.: Man is to computer programmer as woman is to homemaker? Debiasing word embeddings. In: *Advances in Neural Information Processing Systems (NeurIPS)* 29, pp. 4349–4357 (2016).
10. Bender, E.M., Gebru, T., McMillan-Major, A., Shmitchell, S.: On the dangers of stochastic parrots: Can language models be too big? In: *Proceedings of the 2021 ACM Conference on Fairness, Accountability, and Transparency (FAccT '21)*, pp. 610–623. ACM, New York, NY (2021). doi:10.1145/3442188.3445922.
11. Benjamin, R.: *Race After Technology: Abolitionist Tools for the New Jim Code*. Polity Press, Cambridge (2019).
12. Birhane, A., Prabhu, V.U., Kahembwe, E.: Multimodal datasets: misogyny, pornography, and malignant stereotypes. arXiv:2110.01963 [cs.CY] (2021).
13. Crawford, K., Paglen, T.: Excavating AI: the politics of images in machine learning training sets. *AI & Society* 36(4), 1105–1116 (2021). doi:10.1007/s00146-021-01162-8
14. Manovich, L.: *The Language of New Media*. MIT Press, Cambridge, MA (2001).
15. Impett, L.: Digital Art History for Datasets: For an Iconology of AI. In: Brown, K. (ed.) *Artificial Intelligence and Art History: Looking at Images in an Algorithmic Culture*. *Proceedings of the British Academy*, vol. 285, pp. 69–86. The British Academy, London (2026).
16. Rodríguez-Ortega, N.: Ekphrasis Reloaded: Text-to-Image Models and Generative AI. In: Brown, K. (ed.) *Artificial Intelligence and Art History: Looking at Images in an Algorithmic Culture*. *Proceedings of the British Academy*, vol. 285, pp. 87–115. The British Academy, London (2026).
17. Rose, G.: *Visual Methodologies: An Introduction to Researching with Visual Materials*. 5th ed. SAGE, London (2022).
18. Panofsky, E.: *Studies in Iconology: Humanistic Themes in the Art of the Renaissance*. Routledge, London (1972).
19. Zosh, J.M., Hopkins, E.J., Jensen, H., Liu, C., Neale, D., Hirsh-Pasek, K., Solis, S.L., Whitebread, D.: *Learning Through Play: A Review of the Evidence*. The LEGO Foundation (2017). ISBN 978-87-999589-1-7.
20. Toub, T.S., Rajan, V., Golinkoff, R.M., Hirsh-Pasek, K.: Guided play: A solution to the play versus learning dichotomy. In: Geary, D.C., Berch, D.B. (eds.) *Evolutionary Perspectives on Child Development and Education*. Springer, Cham (2016). doi:10.1007/978-3-319-29986-0_5.
21. Salen Tekinbaş, K., Zimmerman, E.: *Rules of Play: Game Design Fundamentals*. MIT Press, Cambridge, MA (2003).
22. Bogost, I.: *Persuasive Games: The Expressive Power of Videogames*. MIT Press, Cambridge, MA (2007).
23. Flanagan, M.: *Critical Play: Radical Game Design*. MIT Press, Cambridge, MA (2009).
24. Shaffer, D.W.: *How Computer Games Help Children Learn*. Palgrave Macmillan, New York, NY (2007).
25. Jenkins, H., Purushotma, R., Weigel, M., Clinton, K., Robison, A.J.: *Confronting the Challenges of Participatory Culture: Media Education for the 21st Century*. MIT Press, Cambridge, MA (2009).
26. Buckingham, D.: *Media Education: Literacy, Learning and Contemporary Culture*. Polity Press, Cambridge (2003).
27. Apperley, T., Beavis, C.: A model for critical games literacy. *E-Learning and Digital Media* 10(1), 1–12 (2013).
28. Kehinde Wiley Studio: Kehinde Wiley, https://kehindewiley.com/, last accessed 2026/02/01.
29. Foka, A.: Experiments in the Relationship between Art History and Text-to-Image Models. In: Brown, K. (ed.) *Artificial Intelligence and Art History: Looking at Images in an Algorithmic Culture*. *Proceedings of the British Academy*, vol. 285, pp. 116–148. The British Academy, London (2026).